\documentclass[twocolumn,superscriptaddress,amsmath, amssymb, amsfonts,preprintnumbers,aps,prd,longbibliography,nofootinbib]{revtex4-1}

\usepackage{scalerel}
\usepackage{color}
\usepackage{amsmath,amsfonts,amssymb}
\usepackage[small,bf,hang]{caption}
\usepackage{slashed}
\usepackage{latexsym,epsfig}
\usepackage{dsfont}
\usepackage{arydshln}
\usepackage{extarrows}
\usepackage{hyperref}
\usepackage{array}

\newcommand{\sfrac}[2]{{\textstyle\frac{#1}{#2}}}

\def\moth{\mathsurround=0pt}
\newdimen\zo \zo=0pt

\def\tick{\leaders\hrule height 0.5ex depth 0pt \hskip 0.5pt}
\def\upboxfill{$\moth \setbox\zo\hbox{\tick}%
  \hskip 3pt\hbox to 0pt{$\tick$\hss}\hrulefill \hbox to 7.5pt{$\tick$\hss}$}

\def\dtick{\leaders\hrule height .34pt depth 0.5ex \hskip 0.5pt}
\def\downboxfill{$\moth \setbox\zo\hbox{\dtick}%
  \hskip 2pt\hbox to 0pt{$\dtick$\hss}\hrulefill \hbox to 2pt{$\dtick$\hss}$}

\def\ov{\bar}

\def\bec{\begin{center}}
\def\ec{\end{center}}

\def\m{\mu}
\def\n{\nu}
\def\r{\rho}

\def\ov{\overline}

\def\nn{\nonumber}

\def\be{\begin{equation}}
\def\ee{\end{equation}}
\def\bea{\begin{eqnarray}}
\def\eea{\end{eqnarray}}
\def\ba{\begin{array}}
\def\ea{\end{array}}

\begin{document}

\title{Aspects of Conformal Gravity and Double Field Theory from a Double Copy Map}

\author{Eric Lescano} 
\email{elescano@irb.hr}
\affiliation{Division of Theoretical Physics, Rudjer Bošković Institute, Bijenička 54, 10000 Zagreb, Croatia}

\author{Gabriel Menezes} 
\email{gabrielmenezes@ufrrj.br}
\affiliation{Departamento de F\'isica, Universidade Federal Rural do Rio de Janeiro, 23897-000, Seropédica, RJ, Brazil}

\author{Jesús A. Rodríguez}
\email{jarodriguez@df.uba.ar}
\affiliation{Universidad de Buenos Aires, FCEyN, Departamento de Física, Ciudad Universitaria, 1428 Buenos Aires, Argentina}


\begin{abstract}

Double Field Theory (DFT) can be constructed as the double copy of a Yang-Mills theory. In this work we extend this statement by including higher-derivative terms. Starting from a four-derivative extension of Yang-Mills whose double copy is known to correspond to a conformal-gravity theory, we obtain a four-derivative theory formulated in double space, which in the pure gravity limit reduces to conformal gravity at quadratic order. This result reveals important aspects for the study of conformal symmetry in the context of DFT through double copy maps.

\end{abstract}

\maketitle

\section{Introduction}

Recently research activities have witnessed a surge of interest in exploring possible relationships between gauge and gravity theories. Such a connection can be instrumental in determining fundamental relations in order to understand more deeply the quantum formulation of gravitational interactions. One successful approach in this direction is the so-called double copy prescription \cite{Dcopy}, which is a map between scattering amplitudes in a broad variety of gravitational and gauge theories (see \cite{Snow} and references therein for an introduction to the subject and its applications). In its original manifestation revealed by Kawai, Lewellen and Tye, closed string tree-level amplitudes can be written in terms of open-string amplitudes~\cite{KLT}. The double-copy formulation also presents correspondences of classical solutions of Yang-Mills theory and gravity~\cite{class} and it has been of paramount importance for scattering-amplitudes methods in the investigation of classical gravitational physics~\cite{class2}.

A very interesting result was recently obtained in \cite{HohmDC}. In this paper the authors demonstrated that the gravitational framework which results from the double copy map of Yang-Mills theory contains a chiral structure, such as the one that happens in the geometry of Double Field Theory (DFT) \cite{Siegel,DFT}. In particular, after applying the double copy procedure to the Yang-Mills theory, the resulting gravitational theory is indeed described by a weak-constrained perturbative DFT, both at quadratic and cubic order, which requires the integration of the generalized dilaton as well as a particular gauge fixing (Siegel's gauge) for describing the cubic interactions, while the quadratic interactions can be described without fixing any gauge.

In DFT a doubled set of coordinates is included. This approach allows for a consistent treatment of both momentum and winding modes in a way that is invariant under T-duality transformations. DFT provides a natural setting for studying the physics of non-geometric backgrounds and has applications in various areas of theoretical physics, including black hole physics, cosmology, and the study of non-commutative geometries \footnote{For reviews see \cite{reviewDFT} and the second lecture of \cite{Electure}.}. Much of the work on DFT has been done on the so-called \textit{strongly constrained} DFT: The inclusion of a T-duality invariant metric as well as a dynamical generalized metric generates a notion of chirality which can be explicitly manifest in terms of projectors. There is an extra fundamental field, the generalized dilaton, which can be used to construct a measure in the double geometry. These fields encode the information related with the massless bosonic fields of the NS-NS sector of string theory, once the antisymmetric field is identified with the Kalb-Ramond field. Pure Yang-Mills theory can be coupled at the DFT level as in \cite{PK,EyS}, but in this work we will consider only gravitational degrees of freedom in the double geometry. 

The double structure of DFT suggest a natural relation with the double copy approach. This was initially explored in \cite{KL} where DFT was used to derive the double copy relationship by showing that the equations of motion of DFT can be mapped onto the equations of motion for Yang-Mills theory. This was done from a perturbative method related with a suitable generalization of the Kerr-Schild ansatz \cite{KS}. This framework can be extended for heterotic DFT \cite{hetKS,AE-HD:21}, KK-DFT \cite{KK} and ExFT \cite{EFT}. The connection between the double copy and the geometry of DFT can also explain properties of the $L_{\infty}$ structure of the latter. It is well known that the DFT Jacobiator is not trivial (but it is given by a trivial parameter) and therefore the algebraic structure of DFT is given by an $L_{\infty}$ algebra with a non-trivial $l_3$ product \cite{HZ}, which measures the failure of the Jacobi identity in the double geometry. The $L_{\infty}$ structure of DFT when the generalized Kerr-Schild ansatz is imposed was studied in \cite{EM}, while the relation between the double copy prescription given in \cite{HohmDC} was recently studied in \cite{HLinfty} generalizing the results of the former.

Based on these previous researches, in this work we go a step further and explore the double copy of higher-derivative theories. In particular we are interested in a result coming from the computation of scattering amplitudes -- double copy of certain higher-derivative Yang-Mills theories corresponds to conformal (super)gravity \cite{DFJ,DFF}. As it was noted in \cite{HohmDC}, double copy requires replacing color factors with a second set of kinematic factors, which come with their own momenta. This substitution leads to a theory in double momentum space or, in position space, a doubled set of coordinates. This suggest a close relation between conformal gravity and a higher-derivative deformation of DFT via double-copy prescription. We start by reviewing the double copy prescription on the Yang-Mills Lagrangian as was done in \cite{HohmDC}, which serves us to present the procedure for the leading order theory (two-derivative case). In the following section we extend the analysis to a Lagrangian containing higher-derivative terms. We find a consistent action on the double space to quadratic order by imposing gauge fixing conditions which reduces to Weyl gravity. 
Then we compute the full cubic order of the theory, which in the pure gravity case contains a larger structure of terms beyond Weyl gravity. Finally we discuss different alternatives to address the inclusion of conformal symmetry in non-perturbative DFT. We work with units such as $\hbar = c = G = 1$.

\section{Double Field Theory as the double copy of Yang-Mills}

As we mentioned previously, it was shown in \cite{HohmDC} that DFT may arise quite naturally from the color-kinematics double copy of Yang-Mills theory. In that work the authors showed that after implementing a double-copy prescription on a Yang-Mills theory the resulting gravitational theory is, at least to cubic order in fields, a week-constrained perturbative DFT which requires the integration of the generalized dilaton. More precisely, at quadratic order they obtained a gauge invariant DFT, whereas the cubic order requires a particular gauge-fixing procedure, the use of the Siegel gauge.

In this section we briefly review the analysis made in Ref.~\cite{HohmDC}. This will help us establish the notation, but mainly the point here is to introduce the double-copy prescription before applying it to a higher-derivative gauge theory related to conformal gravity in the next section. The starting point is a $D$-dimensional Yang-Mills action,
\be
\label{YM_action}
S_{\rm{YM}} = - \frac{1}{4}\int d^{D}x\kappa_{ab}F_{\mu \nu}{}^{a}F^{\mu\nu b}\, ,
\ee
where the Yang-Mills field strength is defined in the standard way
\be
F_{\mu\nu}{}^{a} = 2\partial_{[\mu}A_{\nu]}{}^{a} + g_{\rm{YM}}f^{a}{}_{bc}A_{\mu}{}^{b}A_{\nu}{}^{c}\, ,
\ee
and space-time indices are contracted with a Minkowski metric $\eta_{\mu\nu}={\rm{diag}}(-,+,+,+)$. Passing over to momentum space, the quadratic terms of the gauge action reads (up to a total derivative)
\be
\label{YM_momentum}
S_{\rm{YM}}^{(2)} = - \frac{1}{2}\int_{k}\,\kappa_{ab}\,k^2 \,\Pi^{\mu\nu}(k) A_{\mu}{}^{a}(-k) A_{\nu}{}^{b}(k)\; ,  
\ee
where $\int_{k} \equiv \int d^D k$. The projector $\Pi^{\mu\nu}(k)$ is defined as
\be
\label{PiProjector}
\Pi^{\mu\nu}(k) = \eta^{\mu\nu}-\frac{k^{\mu}k^{\nu}}{k^2}\, , 
\ee 
and obeys the identities  
\be
\label{projIdentity}
\Pi^{\mu\nu}(k)k_{\nu} = 0\;, \quad  \quad \Pi^{\mu\nu}\Pi_{\nu\rho} =  \Pi^{\mu}{}_{\rho}\; . 
\ee

The next step is to take the double copy prescription to construct a gravitational theory. It consists on replacing the color indices by a second set of space-time indices ($a\rightarrow\bar{\mu}$) corresponding to a second set of space-time momenta $\bar{k}^{\bar{\mu}}$. This implies
\be
\label{DcFields}
A_{\mu}{}^{a}(k) \; \longrightarrow \; e_{\mu\bar{\mu}}(k,\bar{k})\;. 
\ee
The Cartan-Killing metric 
$\kappa_{ab}$ must also be substituted following the relation~\cite{HohmDC}
\be
\label{DCCartan}
\kappa_{ab} \; \longrightarrow \; \sfrac{1}{2}\, \bar{\Pi}^{\bar{\mu}\bar{\nu}}(\bar{k})\,, 
 \ee
where the projector $\bar{\Pi}^{\bar{\mu}\bar{\nu}}$ is defined in the same way as the projector $\Pi^{\mu\nu}$ but for barred momenta and indices instead of the original ones. Using these rules the quadratic action \eqref{YM_momentum} becomes,
\be
\label{quadraticDoubleCopy}
S_{\rm DC}^{(2)} = - \frac{1}{4}\int_{k,\bar{k}} k^2\,\Pi^{\mu\nu}(k)  \bar{\Pi}^{\bar{\mu}\bar{\nu}}(\bar{k}) e_{\mu\bar{\mu}}(-k,-\bar{k})e_{\nu\bar{\nu}}(k,\bar{k})\,.  
\ee
It is obvious that the action is symmetric in $k$ and $\bar{k}$ except for the factor $k^{2}$. This asymmetry is solved in DFT due to the so-called level-matching constraint, which states that $\bar{k}^2=k^{2}$.

To complete the construction of the gravitational theory related to \eqref{YM_momentum} we have to Fourier transform the previous action to position space. After expanding the projectors and using the level-matching constraint we find the presence of a non-local term which forces the introduction of an auxiliary scalar field $\phi(k,\bar k)$ such that, when integrated out, the action (\ref{quadraticDoubleCopy}) can be explicitly recovered. It is straightforward  to Fourier transform to a local action in doubled position space:
\begin{align}
\label{localDFTquadPosition}
S_{\rm DC}^{(2)} = \frac{1}{4} & \int d^{D}x \,d^{D}\bar{x}\Big(e^{\mu\bar{\nu}}\square e_{\mu\bar{\nu}}+\partial^{\mu}e_{\mu\bar{\nu}}\,\partial^{\rho}e_{\rho}{}^{\bar{\nu}} \nn \\ 
& + \bar{\partial}^{\bar{\nu}}e_{\mu\bar{\nu}}\,\bar{\partial}^{\bar{\sigma}}e^{\mu}{}_{\bar{\sigma}} - \phi\square\phi + 2\phi\partial^{\mu}\bar{\partial}^{\bar{\nu}}e_{\mu\bar{\nu}}\Big)\, ,
\end{align}
which reproduces the standard quadratic DFT action without the assumption of any gauge choice.  

We have reviewed how to obtain quadratic DFT from the ordinary Yang-Mills action. Next we are going to discuss some relevant aspects of the cubic construction. After Fourier transforming to momentum space the cubic contributions from \eqref{YM_action}, the three-point vertex function arises naturally ($k_{ij} = k_i - k_j$)
\be
\label{Pi_3index}
\pi^{\mu\nu\rho}(k_1,k_2,k_3) = \eta^{\mu\nu} k_{12}^{\rho} 
+ \eta^{\nu\rho} k_{23}^{\mu}
+ \eta^{\rho\mu} k_{31}^{\nu},
\ee
which satisfies the antisymmetric properties of the structure constant. The action at this point can be written as ($A_{i}\equiv A(k_{i})$)
\bea
S^{(3)}_{\rm{YM}} &=& - \sfrac{i g_{\rm{YM}}}{6\left(2\pi\right)^{D/2}}\int_{k_{i}}\delta\left(k_{1}+k_{2}+k_{3}\right)
\nonumber\\
&\times& f_{abc}\pi^{\mu\nu\rho}
A_{1\mu}^{a}A_{2\nu}^{b}A_{3\rho}^{c}\,,
\eea
which shows that an extension of the double copy prescription \eqref{DcFields},\eqref{DCCartan} must be considered in order to include the structure constant. The proper substitution rule is
\be
\label{DC_structure}
f_{abc} \longrightarrow \frac{i}{4} \bar{\pi}^{\bar{\mu}\bar{\nu}\bar{\rho}}\, ,
\ee
defined in the same way as \eqref{Pi_3index} but for barred momentum. One obtains
\bea
\label{cubic}
S_{\rm{DC}}^{(3)} & = & 
\frac{1}{48(2\pi)^{D/2}}\int dK_{1}dK_{2}dK_{3}
\delta(K_{1}+K_{2}+K_{3}) 
\nn \\
&\times& \bar{\pi}^{\bar{\mu}\bar{\nu}\bar{\rho}}
\pi^{\rho\mu\nu}
e_{1\mu\bar{\mu}}e_{2\nu\bar{\nu}}e_{3\rho\bar{\rho}}
\eea
where $K = (k, \bar{k})$, $dK = d^{2D}K$ and $e_{i\mu\bar{\mu}} = e_{\mu\bar{\mu}}(K_i)$. After some manipulations, Fourier transformation to position space and integration by parts, the authors obtain the following cubic action for the double copy of Yang-Mills theory
\bea
\label{YM_dc_cubic}
S^{(3)}_{\rm DC} &=& \frac{1}{8}\int d^{D}x \,d^{D}\bar{x} \ e_{\mu\bar{\mu}}\Big[\,2\partial^{\m}e_{\r\ov\r}\,\ov \partial^{\ov\m}e^{\r\ov\r}-2\partial^{\m}e_{\n\ov\r}\,\ov \partial^{\ov\r}e^{\n\ov\m}\, 
\nn\\ 
&-& 2\partial^{\r}e^{\m\ov\r}\,\ov \partial^{\ov\m}e_{\r\ov\r} +\partial^{\r}e_{\r\ov\r}\,\ov \partial^{\ov\r}e^{\m\ov\m}+\ov \partial_{\ov\r}e^{\m\ov\r}\,\partial_{\r}e^{\r\ov\m}\, \Big].
\eea
It was proven that this action agrees with the cubic DFT action by a gauge fixing condition, integrating out the dilaton. The imposition of a gauge fixing condition is expected considering amplitude computations.

In the next section we are going to apply the double copy map \eqref{DcFields}, \eqref{DCCartan} and \eqref{DC_structure} on the minimal $\left(DF\right)^2$ theory~\cite{DFJ} in order to explore the relation between higher-derivative Double Field Theory and conformal gravity.

\section{Higher-derivative double field theory from the minimal $\left(DF\right)^2$ theory}
 
Our main interest in this work lies in exploring the existence of a relation between conformal gravity and DFT. It was demonstrated that, at the level of amplitudes, the double copy of the following higher-derivative extension of the usual Yang-Mills theory, given by the Lagrangian 
\be
\label{actionhd}
{\cal L} = \frac{1}{2}\kappa_{ab}D_{\mu}F^{\mu\nu a}D_{\rho}F^{\rho}{}_{\nu}{}^{b}\, ,
\ee
corresponds to conformal (super)gravity \cite{DFJ,DFF}. Here, we take this deformation and follow the procedure presented in \cite{HohmDC}. The idea is to explore if this resulting double copy presents a structure that can be interpreted from the double geometry framework.

Considering the gauge covariant derivative defined as
\be
D_{\rho}F_{\mu\nu}{}^{a} = \partial_{\rho}F_{\mu\nu}{}^{a} + g_{\rm{YM}}f^{a}{}_{bc}A_{\rho}{}^{b}F_{\mu\nu}{}^{c}.
\ee
The expansion of the action \eqref{actionhd} up to quadratic terms and its subsequent integration by parts becomes
\be
\label{quadratic_actionhd}
S^{(2)}_{\rm{HD}} = \frac{1}{2}\int d^{D}x\kappa_{ab}\square A^{\mu a}\left(\square A_{\mu}{}^{b} - \partial_{\mu}\partial^{\nu}A_{\nu}{}^{b}\right)\, .
\ee
This expression is interesting because it contains the quadratic expansion of the pure Yang-Mills action \eqref{YM_action}. Going to momentum space it becomes
\be
\label{HD_momentum}
S_{\rm{HD}}^{(2)} = - \frac{1}{2}\int_{k}\,\kappa_{ab}\,k^{4} \,\Pi^{\mu\nu}(k) A_{\mu}{}^{a}(-k) A_{\nu}{}^{b}(k)\; ,  
\ee
After imposing the double copy relations \eqref{DcFields} and \eqref{DCCartan} on the previous action, we obtain
\be
\label{quadraticHD_DoubleCopy}
S_{\rm{HD/DC}}^{(2)} = - \frac{1}{4}\int_{k,\bar{k}} k^4\,\Pi^{\mu\nu}(k)  \bar{\Pi}^{\bar{\mu}\bar{\nu}}(\bar{k}) e_{\mu\bar{\mu}}(-k,-\bar{k})e_{\nu\bar{\nu}}(k,\bar{k}), 
\ee
which, not surprisingly, takes the same form as the Yang-Mills case except for the additional $k^{2}$ contribution. However, this additional contribution avoids the emergence of non-local terms in the action as we can observe after expanding the projectors
\begin{align}
S_{\rm{HD/DC}}^{(2)} = & - \frac{1}{4}\int_{k,\bar{k}}  k^{2}\Big(k^{2} e^{\mu\bar{\nu}}e_{\mu\bar{\nu}} - k^{\mu} k^{\rho} e_{\mu\bar{\nu}}e_{\rho}{}^{\bar{\nu}}  \nn \\ 
& - \bar{k}^{\bar{\nu}}\bar{k}^{\bar{\sigma}} e_{\mu\bar{\nu}}e^{\mu}{}_{\bar{\sigma}} + \frac{1}{k^{2}}k^{\mu}\bar{k}^{\bar{\nu}}k^{\rho}\bar{k}^{\bar{\sigma}}e_{\mu\bar{\nu}}e_{\rho\bar{\sigma}}\Big)\;, 
\end{align}
and hence the introduction of auxiliary fields is not necessary. Transforming the last expression to doubled position space we find the following higher-derivative contributions
\begin{align}
\label{DFDF_DFT}
S_{\rm{HD/DC}}^{(2)} = & - \frac{1}{4}\int d^{D}x d^{D}\bar{x}\Big[\Box e^{\mu\bar{\nu}}\Box e_{\mu\bar{\nu}} - \Box e^{\mu\bar{\nu}}\partial_{\mu}\partial^{\rho}e_{\rho\bar{\nu}}\, \nn \\ 
& - \Box e^{\mu\bar{\nu}}\bar{\partial}_{\bar{\nu}}\bar{\partial}^{\bar{\sigma}}e_{\mu\bar{\sigma}} + \partial^{\mu}\bar{\partial}^{\bar{\nu}}e_{\mu\bar{\nu}}\partial^{\rho}\bar{\partial}^{\bar{\sigma}}e_{\rho\bar{\sigma}}\Big] .
\end{align}
This action can be understood as a higher-derivative extension of DFT with conformal symmetry in the double space (in a classical sense). To clarify this point, we are going to explore the conformal gravity action
\be
\label{Weyl_action}
S_{\rm{CG}} = \int d^{D}x \ \sqrt{-g} \ C_{\mu\nu\rho\lambda}C^{\mu\nu\rho\lambda}\, ,
\ee
with the Weyl tensor given by
\bea
\label{Weyl_tensor}
C_{\mu\nu\rho\lambda} & = & R_{\mu\nu\rho\lambda} - \frac{2}{D-2}\left(g_{\mu[\rho}R_{\lambda]\nu} - g_{\nu[\rho}R_{\lambda]\mu}\right)\, \nn \\
& & + \frac{2}{\left(D-1\right)\left(D-2\right)}Rg_{\mu[\rho}g_{\lambda]\nu}.
\eea
Considering the expansion of $S_{\rm{CG}}$ up to quadratic order for $g_{\mu\nu}=\eta_{\mu\nu} + h_{\mu\nu}$; we obtain
\bea
\label{quadratic_weyl}
S_{\rm{CG}}^{(2)} & = & \frac{D-3}{D-2}\int d^{D}x\left[\left(\Box h^{\mu\nu}\Box h_{\mu\nu} - 2\Box h^{\mu\nu}\partial_{\mu}\partial^{\rho}h_{\rho\nu} \right. \right. \nn \\  
&+& \left. \left. \partial^{\mu}\partial^{\nu}h_{\mu\nu}\partial^{\rho}\partial^{\lambda}h_{\rho\lambda}\right) - \frac{1}{D-1}\left(\Box h - \partial_{\mu}\partial_{\nu}h^{\mu\nu}\right)^{2}\right], 
\nn \\
\eea
where $h=h_{\mu}{}^{\mu}$.

The second term in \eqref{quadratic_weyl} can be removed by imposing the gauge fixing condition $\Box h = \partial_{\mu}\partial_{\nu}h^{\mu\nu}$, related to dilatation symmetry $\delta h_{\mu\nu} = -2\lambda_{D}\eta_{\mu\nu}$. It is straightforward to prove that, after setting $x=\bar{x}$ and properly rescaling the metric (and/or considering a particular volume for the double space when we integrate), in the pure gravity case ($e_{\mu\bar{\nu}}\sim h_{\mu\nu}$) the action \eqref{DFDF_DFT} reduces to \eqref{quadratic_weyl}.

We will now address the double-copy prescription for the cubic terms in \eqref{actionhd}, which can be written as
\bea
\label{cubic_actionhd}
S^{(3)}_{\rm{HD}} & = & g_{\rm{YM}}\int d^{D}x f_{abc}\left[\triangle_{\nu}{}^{a}\left(A_{\rho}{}^{b}\partial^{\rho}A^{\nu c} - A_{\rho}{}^{b}\partial^{\nu}A^{\rho c}\right) \right.\, \nn \\
& & \ \ \ \ \ \ \ \ \ \ \ \ \ \ \ \ \ \ \ \ \ \ \ \  \left. - \partial_{\rho}\triangle_{\nu}{}^{a}A^{\rho b}A^{\nu c}\right]\, ,
\eea
with $\triangle_{\nu}{}^{a} = \Box A_{\nu}{}^{a} - \partial_{\nu}\partial^{\mu}A_{\mu}{}^{a}$.
Following the same logic as in the quadratic case we transform the action to momentum space obtaining
\bea
S_{\rm{HD}}^{(3)} & = & \frac{ig_{\rm{YM}}}{(2\pi)^{D/2}}\int dk_{1}dk_{2}dk_{3}\delta(k_{1}+k_{2}+k_{3})\, \nn \\ 
&\times& f_{abc}k_{1}^{2}\Pi^{\rho\mu}\left(k_{3}^{\nu} - k_{1}^{\nu}\right)A_{1\mu}{}^{a}A_{2\nu}{}^{b}A_{3\rho}{}^{c}, \eea
where $\Pi^{\rho\mu}$ was defined in \eqref{PiProjector}. Before applying the double copy prescription we consider a deformation to the three-point vertex function, given by
\be
\label{Pi3_higher_order}
\Pi^{\mu\nu\rho}(k_{1},k_{2},k_{3}) = \Pi^{\mu\nu}k_{12}^{\rho} + \Pi^{\nu\rho}k_{23}^{\mu} + \Pi^{\rho\mu}k_{31}^{\nu}.
\ee
Notice how similar this is to the standard three-point vertex function defined in Eq.~\eqref{Pi_3index} -- indeed, $\Pi^{\mu\nu\rho}$ can be obtained from $\pi^{\mu\nu\rho}$ by replacing the metric $\eta^{\mu\nu}$ with $\Pi^{\mu\nu}$. This object retains all the properties of the leading-order definition and allows us to write the cubic action as
\be
S_{\rm{HD}}^{(3)} = \frac{ig_{\rm{YM}}}{3(2\pi)^{D/2}}\int_{k_{i}}\delta(k_{i})f_{abc}k_{1}^{2}\Pi^{\mu\nu\rho}A_{1\mu}{}^{a}A_{2\nu}{}^{b}A_{3\rho}{}^{c}\, .
\ee
Analyzing the structure of the action one can realize that the only way to apply the double copy prescription \eqref{DcFields} and \eqref{DCCartan} preserving the symmetry between the coordinates $\mu$ and $\bar \mu$ is to consider $\Pi^{\mu\nu\rho}$ instead of $\pi^{\mu\nu\rho}$ in the substitution rule for the structure constant \eqref{DC_structure}, this is
\be
f_{abc} \longrightarrow \frac{i}{4}\bar{\Pi}^{\bar{\mu}\bar{\nu}\bar{\rho}}\, .
\ee
Then the higher-derivative cubic contributions read,
\bea
\label{cubicHD_DoubleCopy}
S_{\rm{HD}/\rm{DC}}^{(3)} & = & - \frac{1}{24(2\pi)^{D/2}}\int dK_{1}dK_{2}dK_{3}\delta(K_{1}+K_{2}+K_{3})\, \nn \\
&\times& k_{1}^{2}\bar{\Pi}^{\bar{\mu}\bar{\nu}\bar{\rho}}\Pi^{\rho\mu\nu}e_{1\mu\bar{\mu}}e_{2\nu\bar{\nu}}e_{3\rho\bar{\rho}}.
\eea
This expression resembles Eq.~(\ref{cubic}) but encodes the higher-derivative terms of the action \eqref{actionhd}. Expanding this last equation we obtain 
\begin{widetext}
\bea
S_{\rm{HD}/\rm{DC}}^{(3)} & = & - \frac{1}{4(2\pi)^{D/2}}\int dK_{1}dK_{2}dK_{3}\delta(K_{1}+K_{2}+K_{3})\left[\frac{}{}k_{1}^{2}k_{1}^{\rho}\bar{k}_{1}^{\bar{\sigma}}e_{1\mu\bar{\nu}}e_{2}^{\mu\bar{\nu}}e_{3\rho\bar{\sigma}} - k_{1}^{2}k_{2}^{\rho}\bar{k}_{1}^{\bar{\sigma}}e_{1\mu\bar{\nu}}e_{2}^{\mu\bar{\nu}}e_{3\rho\bar{\sigma}}\right.\, \nn \\
& & \ \ \ \ \ \ \ \ \ \ \ \ \ \ \ \ \ \ \ \ \ + k_{1}^{2}k_{2}^{\mu}\bar{k}_{1}^{\bar{\sigma}}e_{1\mu\bar{\nu}}e_{2}^{\rho\bar{\nu}}e_{3\rho\bar{\sigma}} - k_{1}^{2}k_{3}^{\mu}\bar{k}_{1}^{\bar{\sigma}}e_{1\mu\bar{\nu}}e_{2}^{\rho\bar{\nu}}e_{3\rho\bar{\sigma}} - k_{1}^{2}k_{2}^{\mu}\bar{k}_{2}^{\bar{\sigma}}e_{1\mu\bar{\nu}}e_{2}^{\rho\bar{\nu}}e_{3\rho\bar{\sigma}} + k_{1}^{2}k_{3}^{\mu}\bar{k}_{2}^{\bar{\sigma}}e_{1\mu\bar{\nu}}e_{2}^{\rho\bar{\nu}}e_{3\rho\bar{\sigma}}\, \nn \\
& & \ \ \ \ \ \ \ \ \ \ \ \ \ \ \ \ \ \ \ \ \ - k_{1}^{\mu}k_{1}^{\rho}k_{1}^{\lambda}\bar{k}_{1}^{\bar{\sigma}}e_{1\mu\bar{\nu}}e_{2\rho}{}^{\bar{\nu}}e_{3\lambda\bar{\sigma}} + k_{1}^{\mu}k_{1}^{\rho}k_{2}^{\lambda}\bar{k}_{1}^{\bar{\sigma}}e_{1\mu\bar{\nu}}e_{2\rho}{}^{\bar{\nu}}e_{3\lambda\bar{\sigma}} - k_{2}^{\mu}k_{1}^{\rho}k_{1}^{\lambda}\bar{k}_{1}^{\bar{\sigma}}e_{1\mu\bar{\nu}}e_{2\rho}{}^{\bar{\nu}}e_{3\lambda\bar{\sigma}}\, \nn \\
& & \ \ \ \ \ \ \ \ \ \ \ \ \ \ \ \ \ \ \ \ \ + k_{3}^{\mu}k_{1}^{\rho}k_{1}^{\lambda}\bar{k}_{1}^{\bar{\sigma}}e_{1\mu\bar{\nu}}e_{2\rho}{}^{\bar{\nu}}e_{3\lambda\bar{\sigma}} + k_{2}^{\mu}k_{1}^{\rho}k_{1}^{\lambda}\bar{k}_{2}^{\bar{\sigma}}e_{1\mu\bar{\nu}}e_{2\rho}{}^{\bar{\nu}}e_{3\lambda\bar{\sigma}} - k_{3}^{\mu}k_{1}^{\rho}k_{1}^{\lambda}\bar{k}_{2}^{\bar{\sigma}}e_{1\mu\bar{\nu}}e_{2\rho}{}^{\bar{\nu}}e_{3\lambda\bar{\sigma}}\, \nn \\
& & \ \ \ \ \ \ \ \ \ \ \ \ \ \ \ \ \ \ \ \ \ - \bar{k}_{1}^{\bar{\nu}}\bar{k}_{1}^{\bar{\sigma}}\bar{k}_{1}^{\bar{\kappa}}k_{1}^{\rho}e_{1\mu\bar{\nu}}e_{2}^{\mu}{}_{\bar{\sigma}}e_{3\rho\bar{\kappa}} + \bar{k}_{1}^{\bar{\nu}}\bar{k}_{1}^{\bar{\sigma}}\bar{k}_{1}^{\bar{\kappa}}k_{2}^{\rho}e_{1\mu\bar{\nu}}e_{2}^{\mu}{}_{\bar{\sigma}}e_{3\rho\bar{\kappa}} + \bar{k}_{1}^{\bar{\nu}}\bar{k}_{1}^{\bar{\sigma}}\bar{k}_{2}^{\bar{\kappa}}k_{2}^{\mu}e_{1\mu\bar{\nu}}e_{2}^{\rho}{}_{\bar{\sigma}}e_{3\rho\bar{\kappa}}\, \nn \\
& & \ \ \ \ \ \ \ \ \ \ \ \ \ \ \ \ \ \ \ \ \ - \bar{k}_{1}^{\bar{\nu}}\bar{k}_{1}^{\bar{\sigma}}\bar{k}_{2}^{\bar{\kappa}}k_{3}^{\mu}e_{1\mu\bar{\nu}}e_{2}^{\rho}{}_{\bar{\sigma}}e_{3\rho\bar{\kappa}} + \frac{1}{k_{1}^{2}}\bar{k}_{1}^{\bar{\mu}}\bar{k}_{1}^{\bar{\nu}}k_{1}^{\mu}k_{1}^{\nu}\bar{k}_{1}^{\bar{\rho}}k_{1}^{\rho}e_{1\mu\bar{\mu}}e_{2\nu\bar{\nu}}e_{3\rho\bar{\rho}} - \frac{1}{k_{1}^{2}}\bar{k}_{1}^{\bar{\mu}}\bar{k}_{1}^{\bar{\nu}}k_{1}^{\mu}k_{1}^{\nu}\bar{k}_{1}^{\bar{\rho}}k_{2}^{\rho}e_{1\mu\bar{\mu}}e_{2\nu\bar{\nu}}e_{3\rho\bar{\rho}}\, \nn \\
& & \ \ \ \ \ \ \ \ \ \ \ \ \ \ \ \ \ \ \ \ \ \left. + \frac{1}{k_{1}^{2}}\bar{k}_{1}^{\bar{\mu}}\bar{k}_{1}^{\bar{\nu}}k_{1}^{\nu}k_{1}^{\rho}\bar{k}_{2}^{\bar{\rho}}k_{2}^{\mu}e_{1\mu\bar{\mu}}e_{2\nu\bar{\nu}}e_{3\rho\bar{\rho}} - \frac{1}{k_{1}^{2}}\bar{k}_{1}^{\bar{\mu}}\bar{k}_{1}^{\bar{\nu}}k_{1}^{\nu}k_{1}^{\rho}\bar{k}_{2}^{\bar{\rho}}k_{3}^{\mu}e_{1\mu\bar{\mu}}e_{2\nu\bar{\nu}}e_{3\rho\bar{\rho}}\right]\, .
\eea
\end{widetext}
Unlike the quadratic case, the cubic action contains terms that would give rise to non-local terms once we transform to coordinate space, as in the Yang-Mills theory. As we learned from that case, the introduction of an auxiliary scalar field solves this problem. Considering this, the action becomes
\begin{widetext}
\label{Cubic_HD_DFT_momentum}
\bea
S_{\rm{HD}/\rm{DC}}^{(3)} & = & - \frac{1}{4(2\pi)^{D/2}}\int dK_{1}dK_{2}dK_{3}\delta(K_{1}+K_{2}+K_{3})\left[k_{1}^{2}k_{1}^{\rho}\bar{k}_{1}^{\bar{\sigma}}e_{1\mu\bar{\nu}}e_{2}^{\mu\bar{\nu}}e_{3\rho\bar{\sigma}} - k_{1}^{2}k_{2}^{\rho}\bar{k}_{1}^{\bar{\sigma}}e_{1\mu\bar{\nu}}e_{2}^{\mu\bar{\nu}}e_{3\rho\bar{\sigma}}\right.\, \nn \\
& & \ \ \ \ \ \ \ \ \ \ \ \ \ \ \ \ \ \ \ \ \ + k_{1}^{2}k_{2}^{\mu}\bar{k}_{1}^{\bar{\sigma}}e_{1\mu\bar{\nu}}e_{2}^{\rho\bar{\nu}}e_{3\rho\bar{\sigma}} - k_{1}^{2}k_{3}^{\mu}\bar{k}_{1}^{\bar{\sigma}}e_{1\mu\bar{\nu}}e_{2}^{\rho\bar{\nu}}e_{3\rho\bar{\sigma}} - k_{1}^{2}k_{2}^{\mu}\bar{k}_{2}^{\bar{\sigma}}e_{1\mu\bar{\nu}}e_{2}^{\rho\bar{\nu}}e_{3\rho\bar{\sigma}} + k_{1}^{2}k_{3}^{\mu}\bar{k}_{2}^{\bar{\sigma}}e_{1\mu\bar{\nu}}e_{2}^{\rho\bar{\nu}}e_{3\rho\bar{\sigma}}\, \nn \\
& & \ \ \ \ \ \ \ \ \ \ \ \ \ \ \ \ \ \ \ \ \ - k_{1}^{\mu}k_{1}^{\rho}k_{1}^{\lambda}\bar{k}_{1}^{\bar{\sigma}}e_{1\mu\bar{\nu}}e_{2\rho}{}^{\bar{\nu}}e_{3\lambda\bar{\sigma}} + k_{1}^{\mu}k_{1}^{\rho}k_{2}^{\lambda}\bar{k}_{1}^{\bar{\sigma}}e_{1\mu\bar{\nu}}e_{2\rho}{}^{\bar{\nu}}e_{3\lambda\bar{\sigma}} - k_{2}^{\mu}k_{1}^{\rho}k_{1}^{\lambda}\bar{k}_{1}^{\bar{\sigma}}e_{1\mu\bar{\nu}}e_{2\rho}{}^{\bar{\nu}}e_{3\lambda\bar{\sigma}}\, \nn \\
& & \ \ \ \ \ \ \ \ \ \ \ \ \ \ \ \ \ \ \ \ \ + k_{3}^{\mu}k_{1}^{\rho}k_{1}^{\lambda}\bar{k}_{1}^{\bar{\sigma}}e_{1\mu\bar{\nu}}e_{2\rho}{}^{\bar{\nu}}e_{3\lambda\bar{\sigma}} + k_{2}^{\mu}k_{1}^{\rho}k_{1}^{\lambda}\bar{k}_{2}^{\bar{\sigma}}e_{1\mu\bar{\nu}}e_{2\rho}{}^{\bar{\nu}}e_{3\lambda\bar{\sigma}} - k_{3}^{\mu}k_{1}^{\rho}k_{1}^{\lambda}\bar{k}_{2}^{\bar{\sigma}}e_{1\mu\bar{\nu}}e_{2\rho}{}^{\bar{\nu}}e_{3\lambda\bar{\sigma}}\, \nn \\
& & \ \ \ \ \ \ \ \ \ \ \ \ \ \ \ \ \ \ \ \ \ - \bar{k}_{1}^{\bar{\nu}}\bar{k}_{1}^{\bar{\sigma}}\bar{k}_{1}^{\bar{\kappa}}k_{1}^{\rho}e_{1\mu\bar{\nu}}e_{2}^{\mu}{}_{\bar{\sigma}}e_{3\rho\bar{\kappa}} + \bar{k}_{1}^{\bar{\nu}}\bar{k}_{1}^{\bar{\sigma}}\bar{k}_{1}^{\bar{\kappa}}k_{2}^{\rho}e_{1\mu\bar{\nu}}e_{2}^{\mu}{}_{\bar{\sigma}}e_{3\rho\bar{\kappa}} + \bar{k}_{1}^{\bar{\nu}}\bar{k}_{1}^{\bar{\sigma}}\bar{k}_{2}^{\bar{\kappa}}k_{2}^{\mu}e_{1\mu\bar{\nu}}e_{2}^{\rho}{}_{\bar{\sigma}}e_{3\rho\bar{\kappa}}\, \nn \\
& & \ \ \ \ \ \ \ \ \ \ \ \ \ \ \ \ \ \ \ \ \ - \bar{k}_{1}^{\bar{\nu}}\bar{k}_{1}^{\bar{\sigma}}\bar{k}_{2}^{\bar{\kappa}}k_{3}^{\mu}e_{1\mu\bar{\nu}}e_{2}^{\rho}{}_{\bar{\sigma}}e_{3\rho\bar{\kappa}} + k_{1}^{\mu}k_{1}^{\rho}\bar{k}_{1}^{\bar{\nu}}\bar{k}_{1}^{\bar{\sigma}}\phi_{1}e_{2\mu\bar{\nu}}e_{3\rho\bar{\sigma}} - k_{1}^{\mu}k_{2}^{\rho}\bar{k}_{1}^{\bar{\nu}}\bar{k}_{1}^{\bar{\sigma}}\phi_{1}e_{2\mu\bar{\nu}}e_{3\rho\bar{\sigma}}\, \nn \\
& & \ \ \ \ \ \ \ \ \ \ \ \ \ \ \ \ \ \ \ \ \ \left. + k_{2}^{\mu}k_{2}^{\rho}\bar{k}_{1}^{\bar{\nu}}\bar{k}_{1}^{\bar{\sigma}}\phi_{2}e_{1\mu\bar{\nu}}e_{3\rho\bar{\sigma}} - k_{3}^{\mu}k_{2}^{\rho}\bar{k}_{1}^{\bar{\nu}}\bar{k}_{1}^{\bar{\sigma}}\phi_{2}e_{1\mu\bar{\nu}}e_{3\rho\bar{\sigma}}\right]\, .
\eea
The final step in the procedure consists in transforming back to coordinate space. We finally obtain
\bea
\label{Cubic_HD_DFT}
S_{\rm{HD}/\rm{DC}}^{(3)} & = & - \frac{1}{4}\int d^{D}x \ d^{D}\bar{x}\left[\Box\partial^{\rho}\bar{\partial}^{\bar{\sigma}}e_{\mu\bar{\nu}}e^{\mu\bar{\nu}}e_{\rho\bar{\sigma}} - \Box\bar{\partial}^{\bar{\sigma}}e_{\mu\bar{\nu}}\partial^{\rho}e^{\mu\bar{\nu}}e_{\rho\bar{\sigma}} + \Box\bar{\partial}^{\bar{\sigma}}e_{\mu\bar{\nu}}\partial^{\mu}e^{\rho\bar{\nu}}e_{\rho\bar{\sigma}} - \Box\bar{\partial}^{\bar{\sigma}}e_{\mu\bar{\nu}}e^{\rho\bar{\nu}}\partial^{\mu}e_{\rho\bar{\sigma}}\right.\, \nn \\
& & \ \ \ \ \ \ \ \ \ \ \ \ \ \ \ \ - \Box e_{\mu\bar{\nu}}\partial^{\mu}\bar{\partial}^{\bar{\sigma}}e^{\rho\bar{\nu}}e_{\rho\bar{\sigma}} + \Box e_{\mu\bar{\nu}}\bar{\partial}^{\bar{\sigma}}e^{\rho\bar{\nu}}\partial^{\mu}e_{\rho\bar{\sigma}} - \partial^{\mu}\partial^{\rho}\partial^{\lambda}\bar{\partial}^{\bar{\sigma}}e_{\mu\bar{\nu}}e_{\rho}{}^{\bar{\nu}}e_{\lambda\bar{\sigma}} + \partial^{\mu}\partial^{\rho}\bar{\partial}^{\bar{\sigma}}e_{\mu\bar{\nu}}\partial^{\lambda}e_{\rho}{}^{\bar{\nu}}e_{\lambda\bar{\sigma}}\, \nn \\
& & \ \ \ \ \ \ \ \ \ \ \ \ \ \ \ \ - \partial^{\rho}\partial^{\lambda}\bar{\partial}^{\bar{\sigma}}e_{\mu\bar{\nu}}\partial^{\mu}e_{\rho}{}^{\bar{\nu}}e_{\lambda\bar{\sigma}} + \partial^{\rho}\partial^{\lambda}\bar{\partial}^{\bar{\sigma}}e_{\mu\bar{\nu}}e_{\rho}{}^{\bar{\nu}}\partial^{\mu}e_{\lambda\bar{\sigma}} + \partial^{\rho}\partial^{\lambda}e_{\mu\bar{\nu}}\partial^{\mu}\bar{\partial}^{\bar{\sigma}}e_{\rho}{}^{\bar{\nu}}e_{\lambda\bar{\sigma}} - \partial^{\rho}\partial^{\lambda}e_{\mu\bar{\nu}}\bar{\partial}^{\bar{\sigma}}e_{\rho}{}^{\bar{\nu}}\partial^{\mu}e_{\lambda\bar{\sigma}}\, \nn \\
& & \ \ \ \ \ \ \ \ \ \ \ \ \ \ \ \ - \bar{\partial}^{\bar{\nu}}\bar{\partial}^{\bar{\sigma}}\bar{\partial}^{\bar{\kappa}}\partial^{\rho}e_{\mu\bar{\nu}}e^{\mu}{}_{\bar{\sigma}}e_{\rho\bar{\kappa}} + \bar{\partial}^{\bar{\nu}}\bar{\partial}^{\bar{\sigma}}\bar{\partial}^{\bar{\kappa}}e_{\mu\bar{\nu}}\partial^{\rho}e^{\mu}{}_{\bar{\sigma}}e_{\rho\bar{\kappa}} + \bar{\partial}^{\bar{\nu}}\bar{\partial}^{\bar{\sigma}}e_{\mu\bar{\nu}}\bar{\partial}^{\bar{\kappa}}\partial^{\mu}e^{\rho}{}_{\bar{\sigma}}e_{\rho\bar{\kappa}} - \bar{\partial}^{\bar{\nu}}\bar{\partial}^{\bar{\sigma}}e_{\mu\bar{\nu}}\bar{\partial}^{\bar{\kappa}}e^{\rho}{}_{\bar{\sigma}}\partial^{\mu}e_{\rho\bar{\kappa}}\, \nn \\
& & \ \ \ \ \ \ \ \ \ \ \ \ \ \ \ \ \left. + \partial^{\mu}\partial^{\rho}\bar{\partial}^{\bar{\nu}}\bar{\partial}^{\bar{\sigma}}\phi e_{\mu\bar{\nu}}e_{\rho\bar{\sigma}} - \partial^{\mu}\bar{\partial}^{\bar{\nu}}\bar{\partial}^{\bar{\sigma}}\phi\partial^{\rho}e_{\mu\bar{\nu}}e_{\rho\bar{\sigma}} + \partial^{\mu}\partial^{\rho}\phi \bar{\partial}^{\bar{\nu}}\bar{\partial}^{\bar{\sigma}}e_{\mu\bar{\nu}}e_{\rho\bar{\sigma}} - \partial^{\rho}\phi\bar{\partial}^{\bar{\nu}}\bar{\partial}^{\bar{\sigma}}e_{\mu\bar{\nu}}\partial^{\mu}e_{\rho\bar{\sigma}}\right]\, .
\eea    
\end{widetext}
Just as in the quadratic case, we will set $x=\bar{x}$ for the pure gravity case ($e_{\mu\bar{\nu}}\sim h_{\mu\nu}$, $\phi=0$) in order to compare with the cubic contributions coming from \eqref{Weyl_action}. As a first observation we notice that the cubic action obtained from the double copy of the $\left(DF\right)^{2}$ theory contains further contributions apart from the Weyl square action. Some of these contributions vanish after imposing a particular gauge, \textit{e.g.} the harmonic gauge condition, as well as by choosing a particular space-time dimension. It would be interesting to fully understand the physical interpretation of the action \eqref{Cubic_HD_DFT} in terms of symmetry arguments. In the ideal case, one expects a four derivative gauge theory whose double copy map matches with the higher-derivative DFT introduced in \cite{MN}. It is important to take into account that, once we are interested in making contact with a theory coming from a double formalism, it is possible that there are remnants of $O(D,D)$ symmetry in addition to conformal symmetry. Therefore, one possibility is to study this formalism under a generic $D$-dimensional toroidal compactification and to inspect how the $O(D,D)$ multiplets are constructed.

\section{Discussion}

As well known, DFT is defined on a $2D$-dimensional space with coordinates $X_{M}=(\tilde x_{\mu},x^{\mu})$
where $M$ is in the fundamental representation of the $O(D,D)$ group. The field content is the generalized metric ${\cal H}_{M N}$, a symmetric $O(D,D)$ tensor encoding the standard metric $g_{\mu\nu}$ and the Kalb-Ramond field $b_{\mu\nu}$, and the generalized dilaton $d$ which is related with the standard supergravity dilaton $\phi$.

Besides $O(D,D)$ symmetry, the theory is invariant under a generalized notion of diffeomorphism transformations which accounts for usual diffeomorphisms as well as abelian gauge transformations of the $b$-field. Additional symmetries can be incorporated, for instance, the tangent space is enhanced with an extended Lorentz symmetry. Supersymmetry can also be naturally considered. Closure of the generalized diffeomorphisms require the imposition of the so-called \textit{strong constraint} 
\be
\label{strong_const}
\partial_{M}\star\partial^{M}\star = 0\, , \ \qquad \ \partial_{M}\partial^{M}\star = 0\, ,
\ee
where $\star$ denotes $O(D,D)$ fields or combinations of them. 

The theory is equipped with an action principle from which one can derive equations of motion
\be
S_{DFT} = \int d^{D}x d^{D}\tilde{x} \ {\cal R}\left({\cal H},d\right)\, .
\ee
As we can see, the Lagrangian of DFT is given by the generalized Ricci scalar ${\cal R}$, which depends on both the generalized metric and the generalized dilaton,
\bea
\label{scalarDFT}
{\cal R} & = & \frac14 {\cal H}^{MN} \partial_{M}{\cal H}^{KL}\partial_{N}{\cal H}_{KL} - {\cal H}^{MN}\partial_{N}{\cal H}^{KL}\partial_{L}{\cal H}_{MK} \nn \\ && + 8 {\cal H}^{MN} \partial_{M}\partial_{N}d  + 8 \partial_{M}{\cal H}^{MN} \partial_{N}d \nn \\ && - 8 {\cal H}^{MN} \partial_{M}d \partial_{N}d - 2 \partial_{M} \partial_{N} {\cal H}^{MN} \, .
\eea
For a particular solution of the strong constraint, this action reproduces exactly the NS-NS sector of string theory at leading order in $\alpha'$.

At the moment we do not have a conformal formulation of DFT in the sense that it is not clear how to implement conformal symmetry in the double space of the theory. So it is difficult to compare our cubic action~(\ref{Cubic_HD_DFT}) with a given result derived directly from DFT. We speculate this action could be the cubic contribution from an appropriate {\it conformal} DFT defined in a suitable generalized Siegel gauge, a sort of higher-derivative realization of the interpretation given in Ref.~\cite{HohmDC}. Indeed, our results of the previous section clearly suggest the existence of a relation between conformal gravity and some higher-derivative extension of the usual DFT. This scenario is possible due to the symmetric structure between $k$ and $\bar{k}$ that appears in equations \eqref{quadraticHD_DoubleCopy} and \eqref{cubicHD_DoubleCopy}, similar to the ones that emerge in the pure Yang-Mills case. This gives us the opportunity to venture into some intriguing proposals that could account for this relation, which we quickly discuss below.

\begin{itemize}

\item {\it Generalized Kerr-Schild ansatz} -- In Ref.~\cite{KL} a duality invariant analogous of the widely known Kerr-Schild (KS) ansatz was introduced. There, the background generalized metric is linearly and exactly perturbed by a pair of generalized null vectors
\be
\label{KS_ansatz}
{\cal H}_{MN} = \widetilde{{\cal H}}_{MN} + \kappa\left(K_{M}\bar{K}_{N} + K_{N}\bar{K}_{M}\right)\, ,
\ee
with $\kappa$ an arbitrary parameter which quantifies the order of the perturbations and tildes denotes background quantities. Among other conditions, this generalization of the KS ansatz leads to a linearization of the equations of motion of DFT. The parameterization of \eqref{KS_ansatz} leads to 
\bea
g_{\mu\nu} & = & \widetilde{g}_{\mu\nu} + \frac{\kappa}{1+\sfrac{1}{2}\kappa\left(k\cdot\bar{k}\right)}k_{(\mu}\bar{k}_{\nu)}\, \nn \\
b_{\mu\nu} & = & \widetilde{b}_{\mu\nu} - \frac{\kappa}{1+\sfrac{1}{2}\kappa\left(k\cdot\bar{k}\right)}k_{[\mu}\bar{k}_{\nu]}\, .
\eea
Contrary to the ordinary KS ansatz, now the $b$-field can be perturbed but as a consequence, the perturbation of the metric is no longer linear. However, it was proven in the same work that, with this perturbations, the classical double copy structure at the level of the equations of motion can be extended to the entire string NS-NS sector.

In Ref.~\cite{AE-HD:21} the authors extended the analysis to heterotic DFT at order $\alpha'$ and using the classical double copy, they found higher-derivative corrections to the Maxwell equations at order $\kappa$ in the perturbations. Based on this result one of our proposals is to continue this path by extending the order of the perturbations to $\kappa^{2}$ which match with the order of gauge fields expected for the equations of motion coming from \eqref{actionhd}.

\item {\it Conformal symmetry in DFT} -- A different path consists in introducing a new symmetry in the framework of DFT such that Weyl transformations at the supergravity level could be obtained. This seems to be possible with the introduction of an $O(D,D)$ invariant $\Phi$ parameterizing double Weyl transformations.

Consideration of this kind of transformation leads to a particular parameterization of the generalized metric as well as a deformation of the $O(D,D)$ metric. As a consequence, the theory requires a coupling with an extra scalar field to have a conformal invariant action at leading order.

\item {\it Generalized Weyl tensor} -- It is well known that the generalized Riemann tensor of DFT is not fully determined \cite{RiemannDFT} since there are not enough compatibility conditions in the double geometry to fully determine the generalized connection $\Gamma_{MNP}$ in terms of the degrees of freedom of the theory. This makes the direct construction of higher order terms with the structure
\be
{\cal R}_{MNPQ}{\cal R}^{MNPQ}\, \nn
\ee
a difficult task.

This, however, has not been a sufficiently great obstacle to calculate higher-derivative extensions of DFT. A very interesting construction in this sense was put forward in Ref.~\cite{MN}, based on a generalization of the Green-Schwarz mechanism of anomaly cancellation. This theory depends on two parameters whose values are related to different theories: bosonic and heterotic strings and also HSZ theory \cite{HSZ}. Therefore it would be interesting to explore if it is possible to obtain a combination of the parameters such that conformal gravity could be obtained after supergravity reduction.

A different but related approach is to look for the existence of a two-derivative combination of the DFT fields ${\cal C}_{MNPQ}$, mimicking the usual Weyl tensor, such that an action 
\be
{\cal C}_{MNPQ}{\cal C}^{MNPQ}\, \nn
\ee
could be constructed and reduced to conformal gravity.

\end{itemize}

\section{Outlook}

We used the double copy prescription introduced in \cite{HohmDC} as a guiding principle to study the relation between higher-derivative gauge theories and higher-derivative extensions of double field theory (DFT). Exploring the double copy structure of the action (\ref{actionhd}) we obtained promising results for addressing the generalization of conformal symmetry to the double space in which DFT is formulated.

It is known that the double copy of (\ref{actionhd}) is related to a gravity theory with conformal symmetry. As it is possible to express this double copy in a double space formalism, we suggest a relation between conformal gravity and a higher-derivative deformation of DFT. Furthermore, at quadratic order our prescription reduces to Weyl gravity upon enforcing a gauge fixing condition. 

Here we prove that in the pure gravity case, the quadratic action \eqref{DFDF_DFT} precisely agrees with the quadratic graviton terms of conformal gravity. This is expected since these contributions are common to all theories of conformal gravity. The cubic action has additional contributions beyond the Weyl square action. Some of these, but not all, vanishes with specific conditions. The physical interpretation of the action (32) in terms of symmetry arguments is of interest, and different ways of approaching it were proposed.

Finally, based on the relation of the action \eqref{actionhd} and conformal gravity and our construction in double space, we discuss different alternatives to consider conformal symmetry in the context of Double Field Theory. We believe that the proposal discussed in this paper is an important step further in the construction of a non-perturbative conformal double geometry with applications to scattering amplitudes.  

\subsection*{Acknowledgements}
We would like to kindly thank to Felipe Diaz-Jaramillo, Olaf Hohm  and Silvia Nagy for very interesting comments in the first manuscript of this work. EL is supported by the Croatian Science Foundation project IP-2019-04-4168. GM acknowledges the hospitality of the Mani L. Bhaumik Institute for Theoretical Physics at UCLA, where part of this research was carried out. The work of GM has been partially supported by Conselho Nacional de Desenvolvimento Cient\'ifico e Tecnol\'ogico - CNPq under grant 317548/2021-2 and Fundac\~ao Carlos Chagas Filho de Amparo \`a Pesquisa do Estado do Rio de Janeiro - FAPERJ under grants E-26/202.725/2018 and E-26/201.142/2022. The work of JAR is supported by CONICET, and would also like to thank the Hanrieder Foundation for Excellence and the Max Planck Institut für Physik (MPP) for support and hospitality during the early stages of this project.

\end{document}